\documentclass[onecolumn,aps,prb,preprint,floatfix]{revtex4}

\pdfoutput=1

\usepackage{graphicx}
\usepackage{epsfig}
\usepackage{color}
\usepackage{latexsym}
\usepackage{amsmath,amssymb,bm}
\usepackage{natbib}
\usepackage{ulem}

\begin{document}

\title{Ordered Phases and Quantum Criticality in Cubic Heavy Fermion Compounds}

\author{Silke Paschen}
\affiliation{Institute of Solid State Physics, Vienna University of Technology, Wiedner Hauptstr.\ 8-10, 1040 Vienna, Austria}
\author{Julio Larrea J.}
\affiliation{Institute of Solid State Physics, Vienna University of Technology, Wiedner Hauptstr.\ 8-10, 1040 Vienna, Austria}
\affiliation{Physics Department, University of Johannesburg, P.O. Box 524, Auckland Park 2006, South Africa}

\date{\today}

\begin{abstract}
Quantum criticality in cubic heavy fermion compounds remains much less
explored than in quasi-two-dimensional systems. However, such materials are
needed to broadly test the recently suggested global phase diagram for heavy
fermion quantum criticality. Thus, to boost these activities, we review the
field, with focus on Ce-based systems with temperature--magnetic field or
temperature--pressure phase diagrams that may host a quantum critical
point. To date, CeIn$_3$ and Ce$_3$Pd$_{20}$Si$_6$ are the only two among these
compounds where quantum critical behaviour has been systematically investigated.
Interestingly, both show Fermi surface reconstructions as function of the
magnetic field that may be understood in terms of Kondo destruction quantum
criticality.
\end{abstract}


\maketitle

\section{Introduction}

Heavy fermion compounds are at the forefront of research in quantum
criticality.\cite{Ste01.1,Loe07.1} The competition between the Kondo interaction
that tends to screen local moments and the RKKY interaction that favours a
magnetically ordered ground state frequently leads to low-lying phase
transitions. External tuning parameters such as magnetic field or pressure can
either stabilize or weaken the order. This is usually reflected by an
enhancement or a suppression of the ordering temperature. If the ordering
temperature can be fully suppressed to zero, a quantum phase transition is
accessed. The transition between two distinct ground states may happen in a
discontinuous way -- by a first order quantum phase transition, or be
continuous. In the latter case, we refer to it as quantum critical point (QCP).
The interest in these zero-temperature singularities stems from the observation
that finite-temperature properties show unconventional behaviour in the vicinity
of a QCP.

A theoretical description of quantum critical behaviour was derived by extending
the theory of classical criticality to zero temperature.\cite{Her76.1,Mil93.1}
Agreement between experiment and the predicted power laws and scaling relations
was found in some cases, but pronounced deviation have also been observed. These
latter have led to the development of scenarios where the criticality is not
merely due to the vanishing of the order parameter but where another mode is
simultaneously critical. A prominent example is the theory of local quantum
criticality.\cite{Si01.1} In this theory of the antiferromagnetic (AFM) Kondo
lattice with strong two-dimensional spin fluctuations, in addition to the order
parameter also the Kondo interaction is suppressed to zero at the QCP.
Experimental evidence supporting this ``Kondo destruction'' scenario has been
reviewed recently.\,\cite{Si10.2,Si13.1}.

Interestingly, some experiments suggest that both energy scales can also be
detached from each other.\cite{Fri09.1,Cus10.1} This can be rationalized in a
global phase diagram of AFM heavy fermion quantum
criticality.\cite{Si06.1,Col10.2,Si10.1} The two parameters that span this
zero-temperature phase diagram are the Kondo interaction $J_K$ and the
frustration parameter $G$. It was recently pointed out that $G$ may be
identified with spacial dimensionality and, in particular, that cubic compounds
are good candidates for probing the global phase diagram in the limit of ``high
dimensionality'' (low $G$).\cite{Cus12.1}

Here we thus review cubic heavy fermion compounds for which low-lying phase
transitions have been observed. We constrain ourselves to pure (non-substituted)
Ce-based heavy fermion compounds that can be tuned by magnetic field or
pressure. We are not aware of any previous review on this topic and hope that it
will be a useful guide for future research.

\section{Ordered phases}\label{order}

In various cubic Ce-based heavy fermion compounds, low-lying phase transitions
have been observed and were shown to depend on non-thermal tuning parameters
such as magnetic field or pressure. In some of these, the phase transition
temperatures could be fully suppressed to zero (or, more precisely, to a value
below the accessible temperature range). These cases are of interest here as
quantum critical behaviour might emerge from zero-temperature phase transitions.

In Fig.\,\ref{phasediagrams} magnetic field--temperature phase diagrams of
various cubic Ce-based heavy fermion compounds hosting putative quantum critical
points are shown.

\subsection{CeB$_6$}\label{CeB6}

The oldest and most thoroughly studied among these material is certainly
CeB$_6$. It crystallizes in the cubic CaB$_6$-type structure of space group
$Pm\bar 3m$. Its magnetic field--temperature phase diagram\cite{Eff85.1}
contains three different phases: a paramagnetic phase (I) at high temperatures,
an antiferro-quadrupolar (AFQ) phase (II) at intermediate temperatures, and an
antiferromagnetic (AFM) phase (III) at the lowest temperatures
(Fig.\,\ref{phasediagrams}\,a). Under magnetic field, the AFQ ordering
temperature $T_Q$ is enhanced. The nature of the AFM order (phase III or
III$^{\prime}$) depends on the direction along which the magnetic field is
applied. The N\'eel transition $T_N(B)$ is claimed to be of second
order.\cite{Eff85.1} Early evidence for phase II being an AFQ phase was
indirect. NMR experiments (in finite magnetic fields) revealed a splitting of
the $^{11}$B resonance line\cite{Kaw81.1,Tak83.1} below $T_Q$, which is evidence
for AFM order. Also neutron diffraction can only detect the AFM order induced by
a magnetic field.\cite{Eff85.1} Direct evidence for the AFQ phase came from
resonant X-ray scattering (RXS).\cite{Nak01.1} This technique was first used for
$3d$ electron systems.\cite{Mur98.1} In CeB$_6$, the $L_3$ edge of the Ce ion is
probed. The energy level splitting of the $5d$ orbital, induced by the Coulomb
interaction between the $4f$ and the $5d$ orbitals upon orbital ordering, gives
rize to the RXS signal. The ordering wavevector was determined to be
$(1/2,1/2,1/2)$ (in units of $2\pi/a$ where $a$ is the lattice parameter) in
this way.\cite{Mur98.1} High-field measurements up to 60\,T revealed a maximum
of $T_Q(B)$ of almost 10\,K at about 35\,T and a decrease to 8\,K at
60\,T.\cite{Goo04.1} A full suppression of $T_Q$ has, however, not been acheived
yet. Much work has focussed on the determination of the magnetic structure, both
the field-induced one below $T_Q$ and the spontaneous one below
$T_N$.\cite{Eff85.1,Goo04.1,Zah03.1,Pla05.1} Various different ordering wave
vectors have been identified. At first sight such a complex magnetic behaviour
may be surprising in view of the simple crystallographic structure. However, as
will be explained in Sect.\,\ref{CEF} below, it may be attributed to the various
active multipoles.

\subsection{CeTe}\label{CeTe}

CeTe is a member of the much investigated family of Ce monochalcogenides (CeS,
CeSe, CeTe) which crystallize in the faced-centered-cubic NaCl-type
structure and have been intensively investigated for several decades. In zero
magnetic field and at ambient pressure, CeTe undergoes a
second-order\cite{Hul78.1} phase transition at about 2\,K to a type-II AFM
state, with the ordered moments pointing along the magnetic ordering wavevector
$(1/2,1/2,1/2)$ (Ce moments are aligned ferromagnetically within (111) magnetic
planes but point in opposite directions in adjacent (111) planes).\cite{Ott79.1}
The ordered moment was determined to be between $0.15\mu_B$ and
$0.3\mu_B$.\cite{Ott79.1,Rav80.1} The magnetic field--temperature phase diagram
at ambient pressure and at two different hydrostatic pressures is shown in
Fig.\,\ref{phasediagrams}\,b.\cite{Kaw11.1} At ambient pressure, application of
a magnetic field along the crystallographic $[001]$ direction stabilizes a new
phase (phase II) above 1\,T.\cite{Kaw11.1} In another investigation, phase II
was stabilized above 0.5\,T for fields along
$[001]$ and $[110]$ and above 3.5\,T for fields along $[111]$,\cite{Nak04.2}
indicating some sample dependence. Hydrostatic pressure at first stabilizes
phase I, seen by the slight enhancement of the N\'eel temperature to 2.4\,K at
0.45\,GPa and the strongly enhanced critical field $H_c^{I-II}$ of 4\,T at this
pressure.\cite{Kaw11.1} Further increasing pressure weakens phase I again (and
possibly changes its nature, therefore it is referred to as phase I$^{\prime}$)
and stabilizes phase II as the dominating phase at 1.2\,GPa.\cite{Kaw11.1} The
close similarity of the phase diagram of CeTe at 1.2\,GPa and the
ambient-pressure phase diagram of CeB$_6$ has led to the suggestion that phase
II in pressurized CeTe is also an AFQ ordered phase.\cite{Kaw11.1}

\subsection{CeAg}\label{CeAg}

The intermetallic compound CeAg, with the CsCl-type cubic crystal structure at
room temperature, undergoes two successive phase transitions observed in
magnetization,\cite{Mor88.1,Sat96.1} Hall effect,\cite{Sat96.1} magnetic
susceptibility,\cite{Sat96.1} electrical resistivity,\cite{Ihr77.1,Kur83.1} and
specific heat measurements:\cite{Mor88.2} a quadrupolar, tentatively
ferroquadrupolar (FQ),\cite{Mor88.2} transition at $T_{Q} = 17$\,K and a
ferromagnetic (FM) transition at $T_{C} = 5.5$\,K.\cite{Sat96.1}  From
magnetization measurements on CeAg single crystals, which show strong anisotropy
and magnetoelastic hysteresis for all but the $[001]$ direction,\cite{Mor88.1} a
dome-like profile of $T_{Q}(B)$ with a maximum of about 9\,K at 5\,T may be
extracted. However, as no temperature--magnetic field phase diagram appears to
be published, CeAg is not included in Fig.\,\ref{phasediagrams}.

\subsection{CeIn$_3$}\label{CeIn3}

CeIn$_3$ crystallizes in the simple cubic AuCu$_3$-type structure. At 10.2\,K,
it orders antiferromagnetically with an ordering wavevector
$(1/2,1/2,1/2)$\cite{Law80.1} as in CeTe. $^{115}$In NQR measurements revealed
that the moments point along the $\langle 111 \rangle$ direction.\cite{Koh00.1}
The magnetic order was shown to be fully suppressed by a magnetic field of about
60\,T (Fig.\,\ref{phasediagrams}\,c).\cite{Ebi04.1} There is no evidence for
further ordered phases in the accessed temperature and magnetic field range.
Thus, the situation in CeIn$_3$ appears to be simpler than in CeB$_6$, CeTe, and
CeAg.

\subsection{CeOs$_{4}$Sb$_{12}$}\label{CeOs4Sb12}

The filled skutterudite compound CeOs$_{4}$Sb$_{12}$ crystallizes in the
body-centered cubic structure of space group $Im\bar 3$.\cite{Jei77.1} The Ce
atoms are located at the body centre and corners of the cubic structure. They
are surrounded by a cage formed by eight corner-sharing OsSb$_{6}$ octahedra.
The first indication for a low-temperature phase transition in
CeOs$_{4}$Sb$_{12}$ came from specific heat measurements which revealed a
$\lambda$-type anomaly at 1.1\,K. However, as the entropy associated with the
anomaly is only 2\% of $R\ln{2}$ the authors concluded that this phase
transition is extrinsic.\cite{Bau01.1} Subsequent investigations of the specific
heat under magnetic field confirmed the presence of the phase transition and
concluded that it is an intrinsic feature of
CeOs$_{4}$Sb$_{12}$.\cite{Nam03.1,Rot06.1} The evolution of the phase transition
with magnetic field was also tracked by electrical transport\cite{Sug09.1},
NMR/NQR,\cite{Yog09.1} and elastic constant measurements.\cite{Nak07.1} The
magnetic field--temperature phase diagram in
Fig.\,\ref{phasediagrams}\,d\cite{Yog09.1} is delineated by anomalies in the
nuclear spin-lattice relaxation rate $1/T_1$,\cite{Yog09.1} the electrical
resistivity,\cite{Sug09.1} and the specific heat\cite{Nam03.1} of single
crystalline CeOs$_{4}$Sb$_{12}$. The symbols show an initial increase of the
characteristic temperature $T_0$ with increasing magnetic field, and a
suppression of $T_0$ at higher fields. This is similar to the field dependence
of the upper transitions of CeB$_6$ and pressurized CeTe and might thus be due
to quadrupolar order. This is supported by the pronounced softening of the
eleastic constants $C_{11}$ and $C_{44}$ across $T_0$.\cite{Nak07.1}  Evidence
for the shaded area denoted AFM comes from neutron diffraction experiments where
weak AFM reflections with ordering wavevector (1,0,0) were shown to be
suppressed by a magnetic field of 1\,T.\cite{Iwa08.1} Apparently, a signature of
$T_N(\mu_0H)$ is also seen in the electrical resistivity,\cite{Sat03.1} though
no original data are published.

\subsection{Ce$_3$Pd$_{20}$Ge$_6$}\label{Ce3Pd20Ge6}

Ce$_3$Pd$_{20}$Ge$_6$ is a member of the series $R_3$Pd$_{20}$Ge$_6$ ($R$ =
light rare-earth elements) of intermetallic compounds crystallizing in an
ordered variant of the cubic Cr$_{23}$C$_6$-type structure of space group
$Pm\bar 3m$.\cite{Gri94.1} In this structure, the rare-earth atoms occupy two
different crystallographic sites ($4a$ site forming a face-centered cubic
sublattice and $8c$ site forming a simple cubic sublattice), both with cubic
point symmetry ($O_h$ and $T_d$, respectively). In spite of this structurally
more complex situation, the magnetic field--temperature phase diagram of
Ce$_3$Pd$_{20}$Ge$_6$ (Fig.\,\ref{phasediagrams}\,e) is similar to the ones of
CeB$_6$ and pressurized CeTe. The labelling of the different phases was chosen
in analogy with CeB$_6$.\cite{Kit98.1,Nem03.1} In zero field, two successive
phase transitions at 1.2 and 0.7\,K were first revealed by specific heat
measurements.\cite{Kit98.1} The application of a magnetic field along $[001]$
stabilizes the upper transition but suppresses the lower
one.\cite{Kit98.1,Nem03.1} Fields along $[110]$ and $[111]$ lead to more complex
phase diagrams, with phase II being split into two different phases.

Powder neutron diffraction in zero magnetic field revealed AFM order with the
magnetic ordering wavevector $(0,0,1)$ below the lower transition but could not
resolve any magnetic order below the upper transition.\cite{Doe00.1} This
behaviour is distinct from the one observed in a number of other 3-20-6
compounds (Nd$_3$Pd$_{20}$Ge$_6$ and (Nd, Tb, Nd, Ho)$_3$Pd$_{20}$Si$_6$). Here,
below the upper transition the rare earth moments at the $8c$ site order
antiferromagnetically with the ordering wavevector $(1,1,1)$. Below the lower
transition the moments at the $4a$ site order antiferromagnetically with the
ordering wave vector $(0,0,1)$.\cite{Doe00.1a} The absence of the $(1,1,1)$
order in Ce$_3$Pd$_{20}$Ge$_6$,\cite{Doe00.1a} together with the absence of a
clear signature in the magnetic susceptibility at the upper
transition\cite{Kit96.1} but the existence of pronounced minima in the elastic
constants $C_{11}$ and $C_{44}$ at this temperature\cite{Suz99.1} indicate that
this order is quadrupolar in nature. The pronounced softening of the elastic
constant $(C_{11}-C_{12})/2$ and a sizable spontaneous expansion $\Delta L/L =
1.9 \times 10^{-4}$ along the $[001]$ direction were taken as evidence for FQ
order with the order parameter $O_2^0$.\cite{Nem03.1} In an unpublished
diffraction experiment on a Ce$_3$Pd$_{20}$Ge$_6$ single crystal with a slightly
higher N\'eel temperature of 0.75\,K a very weak magnetic signal with the
incommensurate ordering wavevector $(0,0,0.94)$ was observed below 0.45\,K and
suggested to be due to Ce moments at the $8c$ site.\cite{Doe00.1}

\subsection{Ce$_3$Pd$_{20}$Si$_6$}\label{Ce3Pd20Si6}

\begin{figure*}[!ht]
\centerline{\includegraphics[width=18cm,angle=0]{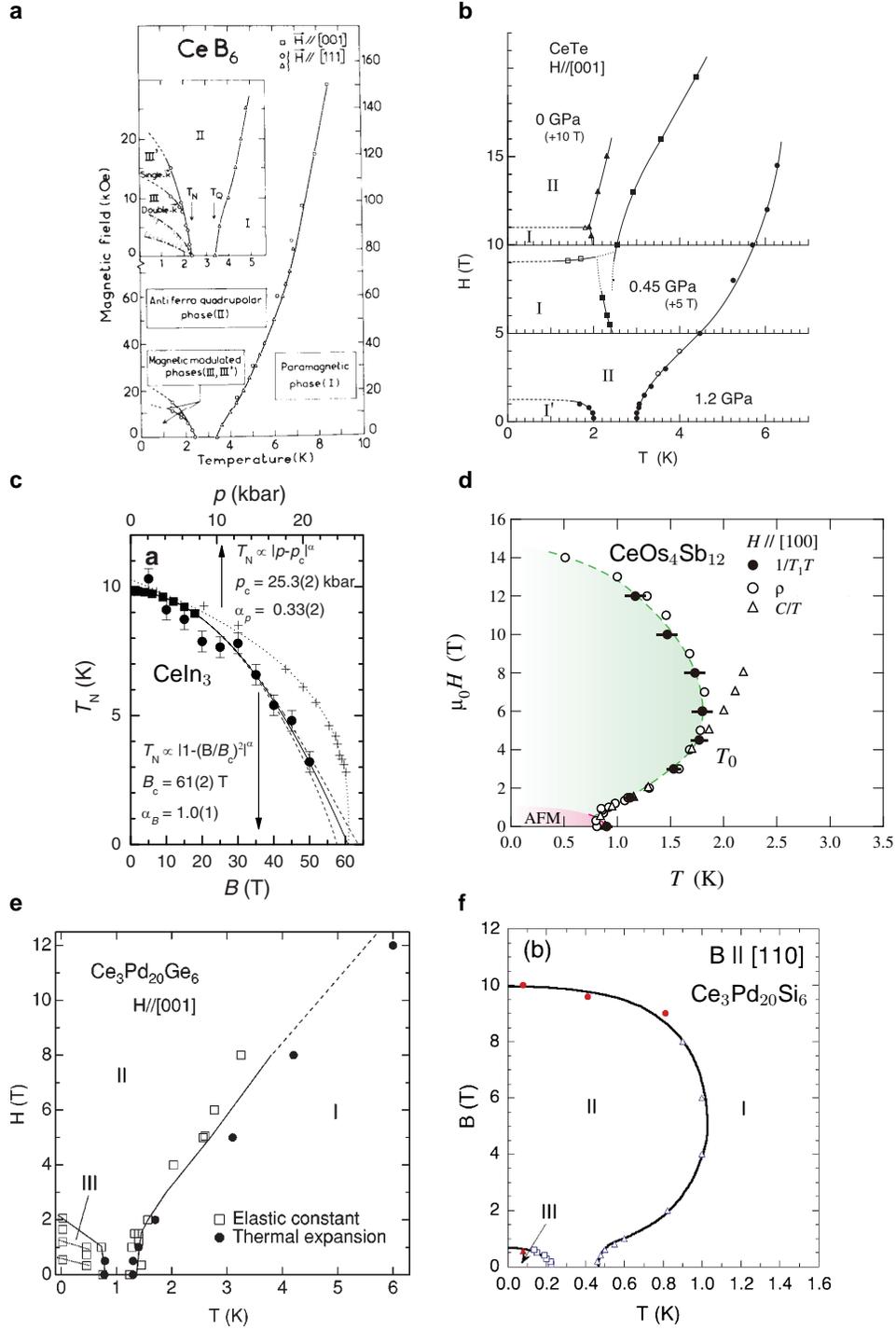}}
\vspace{-2cm}

\caption{(Color online) Magnetic field--temperature phase diagrams of (a)
CeB$_6$,\cite{Eff85.1} (b) CeTe,\cite{Kaw11.1} (c) CeIn$_3$,\cite{Ebi04.1} (d)
CeOs$_4$Sb$_{12}$,\cite{Yog09.1} (e) Ce$_3$Pd$_{20}$Ge$_{6}$,\cite{Nem03.1} and
(f) Ce$_3$Pd$_{20}$Si$_{6}$.\cite{Mit10.1} In panel c, temperature is plotted on
the vertical axis, magnetic field (and pressure, top) on the horizontal axis.}
\label{phasediagrams}
\end{figure*}

Ce$_3$Pd$_{20}$Si$_6$ shows many similarities with its Ge-based sister compound
Ce$_3$Pd$_{20}$Ge$_6$, but also pronounced differences. The
temperature--magnetic field phase diagram for polycrystals was first determined
by specific heat measurements.\cite{Str06.2,Pas07.1,Cus12.1} In zero magnetic
field, Ce$_3$Pd$_{20}$Si$_6$ features two phase transitions at about 0.53\,K and
0.33\,K.\cite{Str06.2} The upper transition is enhanced by a magnetic field to a
maximum value of 1.2\,K at 8\,T.\cite{Cus12.1} The lower transition was later
tracked in detail by transport measurements and was shown to be completely
suppressed by a field of 0.9\,T.\cite{Cus12.1} For single crystals, the
magnetic field--temperature phase diagram was first explored by ultrasound
experiments\cite{Got09.1} and later by magnetization\cite{Mit10.1} and specific
heat measurements.\cite{Ono13.1} Examplarily the phase diagram for field along
$[110]$ is shown in Fig.\,\ref{phasediagrams}\,f.\cite{Mit10.1} Overall, it is
quite similar to the phase diagram obtained for polycrystalline
Ce$_3$Pd$_{20}$Si$_6$.\cite{Cus12.1} For fields along $[001]$ and $[111]$,
however, the upper transition gets sizably modified at fields above
2\,T.\cite{Got09.1,Mit10.1,Ono13.1} For fields along $[001]$, phase II splits
into two phases (II and II$^{\prime}$) which are stable up to about 2\,T and
4\,T, respectively, whereas for fields along $[111]$ phase II is stable up to at
least 14\,T.\cite{Got09.1} Interestingly, the lower transition temperature is
perfectly isotropic with respect to the direction of the magnetic
field.\cite{Mit10.1,Ono13.1} This simple behaviour of phase III is in contrast
to the complex structure, with multiple subphases, and the  anisotropic field
response of phase III in Ce$_3$Pd$_{20}$Ge$_6$.\cite{Sak98.1}

Shortly after the two consecutive phase transitions were revealed by specific
heat measurements\cite{Str06.2} the temperature-dependent magnetic
susceptibility was shown to display a clear anomaly only at the lower of the two
transitions.\cite{Pas07.1} In analogy with Ce$_3$Pd$_{20}$Ge$_6$ this suggests
that the upper transition is quadrupolar in nature. The speculation that
quadrupolar effects are involved was further nourished by the observation of a
pronounced softening of the elastic constants at low temperatures,\cite{Wat07.1}
which is typical of $4f$ electrons with a $\Gamma_8$ ground state. The
temperature and field dependence of the elastic constants were argued to be best
described by AFQ order of $\Gamma_8$ states at the $8c$ site.\cite{Got09.1}

\section{Crystal electric field effects}\label{CEF}

In cubic symmetry, the crystal electric field (CEF) splits the sixfold
degenerate $^2F_{5/2}$ multiplet of the Ce$^{3+}$ $4f^1$ state into a $\Gamma_7$
doublet and a $\Gamma_8$ quartet. The $\Gamma_7$ system supports only dipoles
but in the $\Gamma_8$ system there are various active multipoles: three dipoles,
five quadrupoles, and seven octupoles.\cite{Shi97.1} Due to the different
symmetries of the quadrupolar moments and the dipolar and octupolar moments,
quadrupolar ordering cannot induce additional moments in zero magnetic field.
However, the application of a field lowers the symmetry and thus quadrupolar
order will in general be accompanied by other moments induced by the external
field.\cite{Shi97.1}

In heavy fermion systems, the Ce$^{3+}$ $4f^1$ state interacts with conduction
electrons ($c$--$f$ hybridization). This leads to a broadening of the CEF split
levels. A strong and anisotropic $c$--$f$ hybridization might, in addition, even lead
to a lowering of the cubic point symmetry and thus to a modification of the
theoretically expected level scheme. Thus, the experimental determination of the
level scheme in heavy fermion systems is far more challenging than in well
localized $f$ electron systems.

Generally, the energy splitting between the ground state and excited levels can
be measured by any spectroscopic technique, provided the matrix element for that
transition is sufficiently large. Typically, inelastic neutron scattering is
employed,\cite{Zir84.1,Ros85.1,Kel99.1,Adr07.1,Dee10.1} but Raman
spectroscopy\cite{Zir84.1} has also proven useful. If the CEF splitting is small
and the broadening due to the $c$--$f$ hybridization is large, it may be
difficult to distinguish the CEF excitation (e.g., inelastic $\Gamma_7
\rightarrow \Gamma_8$ transition) from quasielastic scattering (e.g., elastic
$\Gamma_7 \rightarrow \Gamma_7$ scattering).\cite{Law80.1} To decide which of
the levels is the ground state requires a careful determination of the
scattering intensities/magnetic form factors and usually further experimental
evidence (entropy, magnetization, elastic constants, ...) is used to support the
assignment. A rather new technique that can provide direct evidence for the CEF
ground state level is resonant inelastic X-ray scattering.\cite{Wil10.1}
However, for systems with cubic point symmetry information on the ground state
wave function has to date only been extracted in finite magnetic
fields.\cite{Wil11.1}

In CeB$_6$, the situation has long been unclear due to conflicting level schemes
proposed from various thermal, magnetic and elastic data. Finally, inelastic
neutron scattering revealed an energy splitting of 535\,K (46\,meV). From the
temperature-dependent frequency shift of the Raman scattering signal a
splitting of 30\,K was deduced and taken as evidence for the (weakly split)
$\Gamma_8$ quartet being the ground state.\cite{Zir84.1} This is consistent with
the experimentally observed AFQ ordering in zero field and the field-induced
magnetic order below $T_Q(B)$.

In CeTe, on the other hand, the $\Gamma_7$ doublet was reported to be the ground
state and the $\Gamma_8$ quartet the excited state at 32\,K above.\cite{Ros85.1}
Unfortunately, both the original neutron scattering data supporting the energy
splitting of 32\,K as well as the high-field magnetization measurements
supporting the $\Gamma_7$ ground state appear to be unpublished. Specific heat
and magnetic susceptibility failed to give clear indications for the CEF
splitting\cite{Hul78.1} but elastic constants measurements were found to be
consistent with the splitting of 32\,K.\cite{Mat88.1} Quadrupolar order cannot
directly result from a $\Gamma_7$ ground state because it does not have
quadrupolar degrees of freedom. However, a mixing of $\Gamma_8$ components into
the $\Gamma_7$ ground state can occur\cite{Han84.1} and has been proposed as
mechanism for the pressure induced putative AFQ order of CeTe.\cite{Kaw11.1} In
fact, fits to the temperature dependence of the magnetization at various
pressures yielded a reduction of the CEF splitting with pressure. This was
suggested to be responsible for stabilizing the AFQ order under
pressure.\cite{Kaw11.1} Also the enhancement of the AFQ phase transition
temperature under magnetic field is consistent with this scenario.\cite{Han84.1}

In CeAg, inelastic neutron scattering experiment observed a CEF excitation at
about 265\,K (23\,meV).\cite{Sch78.1} Magnetoelastic measurements assigned the
$\Gamma_{8}$ quartet as the CEF ground state.\cite{Tak81.1,Mor88.1}

The CEF level scheme of CeIn$_3$ was explored by inelastic neutron scattering on
powder\cite{Law80.1,Mur93.1} and single crystalline samples.\cite{Kna03.1} A 
broadened CEF excitation is observed at about 140\,K
(12\,meV).\cite{Law80.1,Mur93.1,Kna03.1} From an analysis of the magnetic
susceptibility the $\Gamma_7$ doublet was suggested to be the ground
state.\cite{Bus69.1} This was later confirmed by magnetic form factor
measurements which are best explainned with a $\Gamma_7$ ground state with some
admixture of the $\Gamma_8$ wavefunction.\cite{Bou83.1}

In CeOs$_{4}$Sb$_{12}$, the CEF level scheme is still controversial. Inelastic
neutron scattering on CeOs$_{4}$Sb$_{12}$ powder revealed two broad magnetic
excitations centered at 315\,K (27\,meV) and 555\,K (48\,meV).\cite{Adr07.1} One of them
might be due to a CEF excitation, broadened by the $c$--$f$ hybridization, even
though an interpretation of these features as indirect and direct transition
across a hybridization gap was preferred.\cite{Adr07.1} The temperature
dependence of the magnetic susceptibility is best described by a level scheme
with a $\Gamma_7$ ground state separated by 325\,K (28\,meV) from the
$\Gamma_8$ excited state.\cite{Bau01.1} The exponential temperature dependence
of the NQR relaxation rate $1/T_1$ with an activation energy of
330\,K\cite{Yog05.1} may be taken as confirmation of this splitting. However,
the authors take the lack of a minimum of the elastic constant
$(C_{11}-C_{12})/2$,\cite{Nak07.1} expected for the above level scheme at
$0.5\times 330$\,K, as evidence against this interpretation and suggest instead that a
hybridization gap opens in CeOs$_{4}$Sb$_{12}$.\cite{Yog05.1} 

Inelastic neutron scattering on Ce$_3$Pd$_{20}$Ge$_{6}$ revealed
excitations at 60\,K (5.2\,meV) and 46\,K (4\,meV).\cite{Kel99.1} Since the
magnetic entropy reaches $R \ln{4}$ per Ce-mole at about 10\,K the $\Gamma_8$
wave function was assumed to be the ground state.\cite{Kit96.1} From the ratio
of the intensities of the two excitations the one at 60\,K was associated with
the $4a$ site and the one at 46\,K  with the $8c$ site.\cite{Kel99.1}

In inelastic neutron scattering on Ce$_3$Pd$_{20}$Si$_{6}$, on the other hand,
only one excitation at 44\,K (3.8\,meV) was clearly
resolved.\cite{Pas08.1,Dee10.1} This could either imply that the excitation
energy is similar for both Ce sites and that the peak thus contains both
excitations, or that the excitation energy of the second site is much smaller or
much larger than 44\,K. An analysis with a very low excitation energy of only
3.6\,K (0.31\,meV) for the $8c$ cite was attempted\cite{Dee10.1} but the
subtraction of the large quasielastic signal puts large uncertainties on this
analysis. The magnetic entropy exceeds $3R\ln{2}$ per formula unit already at
1.5\,K.\cite{Tak95.1} If the excitation energy is 44\,K for both Ce sites, this
would imply that at least at one of the Ce sites assumes a $\Gamma_8$ ground
state. Further investigations are needed to clarify the situation.

\section{Structural transitions}

One important question is whether the orbital ordering observed in most of the
systems discussed above is associated with a structural transition. This may
reduce the cubic symmetry of the system and thus change the degree of
frustration (the value of $G$ in the global phase diagram). In transition metal
oxides, orbital order frequently leads to a Jahn-Teller lattice distortion and
thus to a lowering of the symmetry.\cite{Kaj99.1,Lin00.1} This is due to the
large spacial spread of the $3d$ electron orbitals which leads to a strong
coupling with the lattice. $4f$ electrons are more localized and thus their
coupling with atomic displacements is expected to be weaker.

In CeB$_6$ a high-resolution neutron powder diffraction investigation has
searched for lattice distortions in the vicinity of $T_Q$ and $T_N$. Within the
accuracy of the experiment, which was 0.0003\,\AA\ for the lattice parameter and
$0.003^{\circ}$ for the angle, no distortion cound be detected.\cite{Zah03.1}

In CeTe under ambient conditions in magnetic field and pressure, an X-ray
diffraction study could not resolve any structural distortion at the transition
into phase I. The upper limit for the amplitude of possible distortions was set
to $10^{-3}$ lattice units by this experiment.\cite{Ott79.1}  Only at very high
pressures (8\,GPa), a structural phase transition from the NaCl-type structure
to the CsCl-type structure (both cubic) occurs.\cite{Leg83.1}

Neutron diffraction experiments on CeAg revealed that the transition at the
putative FQ transition $T_{Q}$ is accompanied by a sizable lattice distortion
($c/a - 1 = 1.9$\,\%) and thus by a lowering of the symmetry from cubic to
tetragonal.\cite{Sch78.1}

In the temperature--magnetic field phase diagram of CeIn$_3$ the N\'eel
transition is the only phase transition. Thus, any structural distortion would
be expected to accompany this transition. However, neutron diffraction could not
reveal any structural change at $T_N$ and the magnetic reflections can be
indexed on a doubled unit cell of cubic symmetry.\cite{Law80.1}

For CeOs$_{4}$Sb$_{12}$, detailed structural investigations across $T_0$ are not
yet available.

In Ce$_3$Pd$_{20}$Ge$_6$, a spontaneous expansion by $\Delta L/L = 1.9 \times
10^{-4}$ along the $[001]$ direction, associated with a transition from cubic to
tetragonal symmetry, was observed at the transition between phase I and II. It
was taken as evidence for $O_2^0$-type FQ ordering.\cite{Nem03.1}

On Ce$_3$Pd$_{20}$Si$_6$, high-resolution neutron diffraction experiments
($\Delta d/d \approx 0.0025$) could not resolve any structural distortion
between 40\,mK and room temperature.\cite{Dee10.1} The perfectly isotropic
behaviour of phase III further supports the absence of a symmetry lowering
structural transition.\cite{Cus12.1}

\section{Continuous vs first order transitions}

For a finite-temperature phase transition to lead to quantum criticality it
needs to remain continuous as it is suppressed down to $T=0$. A much employed
technique to reveal the order of a finite-temperature phase transition is to
search for thermal hysteresis effects. The presence of a thermal hysteresis
across the phase transition temperature revealed in any of the physical
properties is a strong indication for the first order nature of the transition.

Among the compounds discussed above, only for CeOs$_{4}$Sb$_{12}$ pronounced
thermal hysteresis effects were revealed. The temperature dependence of the
nuclear spin lattice relaxation rate as well as the full width at half maximum
of the NQR spectrum of the Sb nuclei show a hysteresis below $T_{0}$ upon
heating and cooling the sample.\cite{Yog05.1} This indicates that the phase
transition at $T_{0}$ is of first order.

\section{Effect of pressure}

\begin{figure}
\centerline{\includegraphics[width=17cm,angle=0]{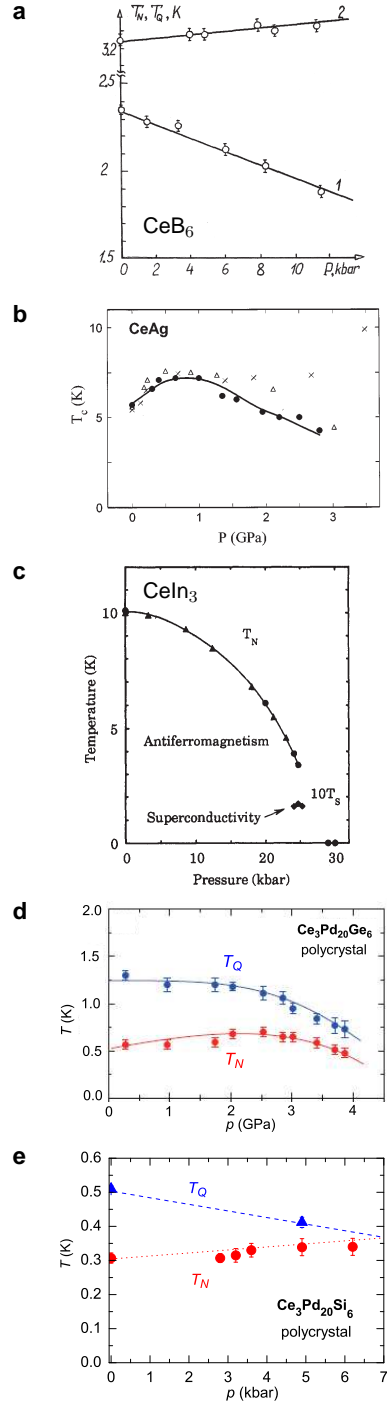}}
\vspace{-2cm}

\caption{(Color online) Temperature--pressure phase diagrams of (a) CeB$_{6}$,\cite{Bra85.1}
(b) CeAg ($\bullet$ from $\chi(T)$,\cite{Cor97.1} $\bigtriangleup$ from $\rho(T)$,\cite{Eil81.1} and $\times$ from from $\rho(T)$\cite{Fuj87.1}), (c)
CeIn$_{3}$,\cite{Mat98.1} (d) Ce$_{3}$Pd$_{20}$Ge$_{6}$,\cite{Hid12.1} and (e)
Ce$_{3}$Pd$_{20}$Si$_{6}$.\cite{Lar13.1} $T_{N}$, $T_C$, $T_Q$, and $T_S$ denote the N\'eel, Curie, quadrupolar, and superconducting transition temperatures. Measurements were performed on single crystals if not stated otherwise.}
\label{pressure}
\end{figure}

In Sect.\,\ref{order} we have discussed the magnetic field--temperature phase
diagrams of various cubic Ce-based heavy fermion compounds. Some of these have
also been studied under pressure. The purpose of this section is to review these
pressure studies and to highlight cases where a pressure-tuned quantum critical
point (QCP) has been accessed or may be in reach. In Fig.\,\ref{pressure} the
temperature--pressure phase diagrams of five different cubic heavy fermion
compounds are shown.

For single crystalline CeB$_6$, the phase diagram under hydrostatic pressure
(Fig.\,\ref{pressure}\,a) was determined by magnetic susceptibility
measurements in small fields applied along the $[100]$ direction, using
extrapolations to zero field.\cite{Bra85.1} The decrease of $T_{N}$ and
the increase of $T_{Q}$ with pressure were later confirmed by magnetization
measurements in larger fields along $[110]$\cite{Uwa00.1} and
$[100]$.\cite{Ike07.1} The electrical resistivity $\rho(T)$ of single
crystalline CeB$_{6}$ was measured under hydrostatic pressure up to
130\,kbar.\cite{Kob00.1} A gradual increase of $T_{Q}$ with pressure is observed
up to 50\,kbar but at larger pressures the characteristic feature in $\rho(T)$
associated with the transition gets lost. Unfortunately, in this experiment
$\rho(T)$ could not detect the transition at $T_{N}$. Thus, it is still open
whether a pressure-induced QCP can be reached in CeB$_6$.

In CeAg, both the putative FQ transition at $T_{Q}$ and the FM transition at
$T_C$ have been studied under pressure. $T_{Q}(p)$ was tracked by electrical
resistivity measurements. Below 2.2\,kbar, $T_{Q}$ (referred to as $T_{M1}$ by
the authors\cite{Kur83.1}) is independent of pressure and no thermal hysteresis
is observed. At 2.2\,kbar and above, $T_{Q}$ is strongly enhanced up to room
temperature at 20\,kbar and shows pronounced thermal
hysteresis\cite{Kur83.1,Fuj87.1}. $T_{C}(p)$ (Fig.~\ref{pressure}\,b) was
derived from magnetic susceptibility measurements under hydrostatic
pressure\cite{Cor97.1} and from electrical resistivity measurements under
hydrostatic\cite{Fuj87.1} and quasihydrostatic\cite{Eil81.1} pressure. Below
10\,kbar, according to all three experiments, $T_C$ increases with pressure. At
higher pressures, $T_C$ decreases according to the former two experiments but
saturates and then further increases according to the latter experiment.  Sample
dependencies below $T_Q$ were attributed to different initial strains in the
samples due to different growth processes\cite{Mor88.2}. Such effects might be
responsible for these conflicting results. At pressures above 30\,kbar no signs
of a FM transition were detected down to 2\,K.\cite{Cor97.1} This might suggest
that ferromagnetism is suppressed in a sharp first order transition at pressures
slightly above 30\,kbar. Clearly, further experiments are needed to clarify the
situation.

In CeIn$_{3}$, the pressure evolution of the N\'eel temperature was first
determined by electrical resistivity measurements, which show a discontinuity in
the temperature gradient at $T_{N}$.\cite{Wal97.1} Hydrostatic pressure
continuously suppresses $T_{N}$ from 10\,K at ambient pressure to below 3\,K at
25\,kbar. Beyond this pressure, the signature in the electrical resistivity gets
lost. A superconducting dome appears in the vicinity of the critical pressure
for the full suppression of $T_N$, estimated to about 26\,kbar
(Fig.\,\ref{pressure}\,c).\cite{Wal97.1} The pressure dependence of
$T_N$\cite{Mat98.1} is plotted together with the field dependence of
$T_N$\cite{Ebi04.1} in Fig.\,\ref{phasediagrams}\,c.

Recently, electrical resistivity under pressure was measured on polycrystalline
samples of Ce$_{3}$Pd$_{20}$Ge$_{6}$.\cite{Hid12.1}. At ambient pressure, the
temperature dependence of the electrical resistivity shows a clear kink at
$T_{N}$ and a broader feature at $T_{Q}$. The evolution of both features with
pressure is shown in Fig.\,\ref{pressure}\,d.\cite{Hid12.1} Most interesting in
the context of quantum criticality is that, at high pressures, both $T_{N}$ and
$T_{Q}$ decrease. A linear extrapolation of the data above 30\,kbar leads to an
estimation of the critical pressures $p_{Q,c}=75$\,kbar and $p_{N,c} = 69$\,kbar
for the full suppression of $T_{Q}$ and $T_{N}$, respectively. Uniaxial pressure
investigations on single crystalline  Ce$_{3}$Pd$_{20}$Ge$_{6}$ up to 3\,kbar
could not resolve any change of $T_{Q}$ and $T_N$, in agreement with the above
results. However, sizable changes are seen in finite magnetic
fields.\cite{Yam02.1}

Polycrystalline Ce$_{3}$Pd$_{20}$Si$_{6}$ has only very recently been studied
under pressure. The evolution of $T_N$ and $T_Q$ with hydrostatic pressure
(Fig.\,\ref{pressure}\,e) was determined by  electrical resistivity,
magnetoresistance and specific heat measurements. \cite{Lar13.1} $T_N$ increases
almost linearly with pressure whereas $T_Q$ decreases. A linear extrapolation of
both dependencies suggests that $T_N(p)$ and $T_Q(p)$ intersect at 7\,kbar.
Ongoing measurements under higher pressure will reveal whether $T_Q$ and/or
$T_N$ can be fully suppressed by pressure.

\section{Quantum critical behaviour}

In spite of the wealth of information on magnetic field- and pressure-tuned
phase transitions in cubic Ce-based heavy fermion compounds presented above,
investigations of quantum criticality have been carried out only in few cases.

\begin{figure}
\centerline{\includegraphics[width=22cm,angle=0]{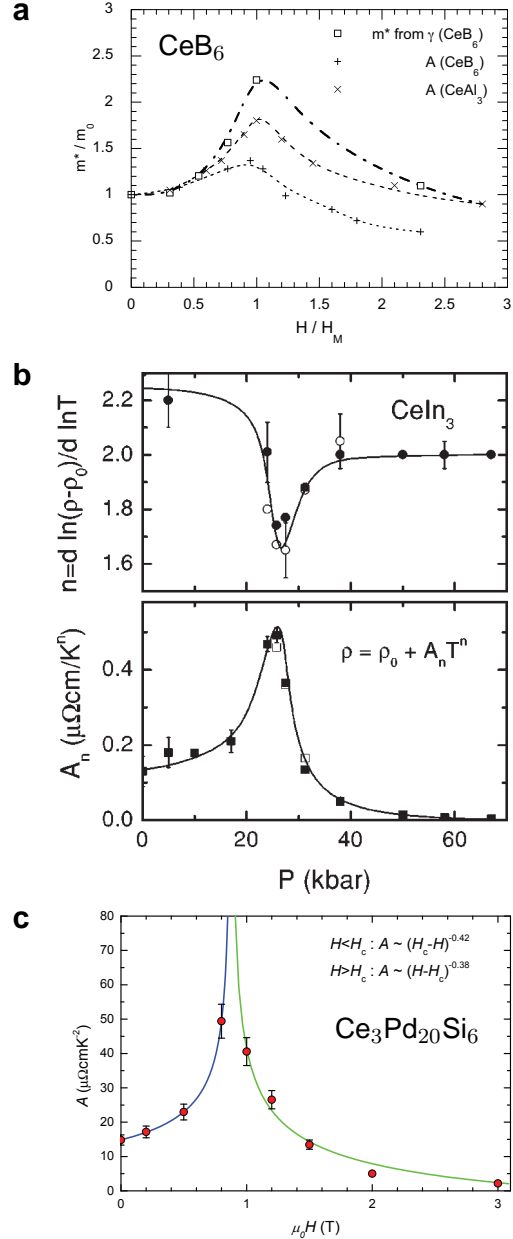}}
\vspace{-9cm}

\caption{(Color online) Characteristics of quantum criticality in CeB$_6$, CeIn$_3$, and
Ce$_3$Pd$_{20}$Si$_6$. (a) Reduced effective mass deduced from $\gamma$ and $A$
vs reduced magnetic field ($\mu_0 H_M = 1.3$\,T) for CeB$_6$,\cite{Mar88.1} (b) resistivity exponent
$n$ and generalized resistivity coefficient $A_n$ vs pressure for
CeIn$_3$,\cite{Kne01.1} and (c) $A$ vs magnetic field for
Ce$_3$Pd$_{20}$Si$_6$.\cite{Lor12.3} The plot in (a) was remade by C.\ Marcenat
because the corresponding plot in the original reference\cite{Mar88.1} is not electronically available is sufficient quality.}
\label{QCP}
\end{figure}

An interesting early observation on CeB$_6$ is that the Sommerfeld coefficient
$\gamma$ of the Fermi liquid (FL) contribution to the specific heat, $C = \gamma
T$, as well as the $A$ coefficient of the FL form of the electrical resistivity,
$\rho = \rho_0 + A T^2$, show cusp-like enhancements near the critical field for
the suppression of the AFM phase of about 1.3\,T
(Fig.\,\ref{QCP}\,a).\cite{Mar88.1} Such field dependences are today considered
as strong indications for quantum criticality.

CeIn$_3$ is certainly the most prominent Ce-based quantum critical compound with
cubic crystal structure.\cite{Wal97.1,Mat98.1,Kne01.1,Ebi04.1,Seb09.1} As
discussed above, the N\'eel transition can be continuously suppressed to zero by
either a magnetic field or by pressure. In the vicinity of the critical pressure
$p_c = 26.5$\,kbar, the temperature dependence of the electrical resistivity
shows clear deviations from FL behaviour and is better described by the
non-Fermi liquid (NFL) form $\rho = \rho_0 + A_n T^n$
(Fig.\,\ref{QCP}\,b).\cite{Kne01.1}

Also in Ce$_3$Pd$_{20}$Si$_6$, quantum critical behaviour has been
observed,\cite{Cus12.1} with the electrical resistivity close to the critical
field $\mu_0H_c =0.9$\,T being best described by a NFL form with an exponent
$n=1$.\cite{Pas07.1} The $A$ coefficient of the FL form, determined at the
lowest temperatures away from $H_c$, is strongly enhanced towards $H_c$
(Fig.\,\ref{QCP}\,c).\cite{Lor12.3} Isothermal field dependences of the
longitudinal and transverse magnetoresistance and the Hall coefficient show a
crossover at a characteristic field $H^{\ast}$ that coincides with $H_c$ only in
the zero-temperature limit.\cite{Cus12.1} The width of the crossover sharpens
with decreasing temperature in a pure power-law fashion, resulting in a sharp
step in the extrapolation to zero temperature, much like the situation in the
tetragonal heavy fermion compound YbRh$_2$Si$_2$.\cite{Pas04.1,Fri10.2} This
extrapolated zero-temperature discontinuity in the transport properties was
interpreted\cite{Pas04.1,Fri10.2,Cus12.1} as Kondo
destruction.\cite{Si01.1,Col01.1} In simple terms, at a Kondo destruction QCP
the $f$ component of the itinerant electrons localizes and thus drops out of the
Fermi sea. This changes the Fermi volume from ``large'' to ``small''. As a
consequence, local moment order may occur. As low-dimensional spin fluctuations
are needed in the formulation of the theory,\cite{Si01.1} it was considered
unlikely that a Kondo destruction QCP could appear in three-dimensional
systems. In cubic Ce$_3$Pd$_{20}$Si$_6$, the AFM phase is
isotropic\cite{Mit10.1,Ono13.1} and thus it seems plausible to consider it as
three-dimensional system. The global phase diagram for AFM heavy fermion
compounds\cite{Si06.1,Si10.1,Cus10.1,Col10.2} provides a way to think about the
unexpected Kondo destruction in this compound.\cite{Cus12.1} In the region of
small values of the magnetic frustration parameter $G$, a transition from an
antiferromagnet with small Fermi surface to an antiferromagnet with large Fermi
surface is predicted. In fact, the Kondo destruction QCP in
Ce$_3$Pd$_{20}$Si$_6$ occurs at the critical magnetic field for the suppression
of antiferromagnetism (phase III in Fig.\,\ref{phasediagrams}\,f), inside the
putative AFQ ordered phase (phase II in Fig.\,\ref{phasediagrams}\,f). Magnetic
fields are likely to induce magnetic dipolar moments on top of the ordered
quadrupolar moments.\cite{Cus12.2} Thus, it appears plausible that the Kondo
destruction QCP in Ce$_3$Pd$_{20}$Si$_6$ indeed separates to antiferromagnetic
phases.

Interestingly, in CeIn$_3$ the effective mass of heavy r-orbits observed in de
Haas-van Alphen (dHvA) experiments appears to diverge at a field of about 40\,T
that is well below the critical field of 61\,T for the suppression of
antiferromagnetism.\cite{Seb09.1} The AFM order is claimed to change its nature
from itinerant to local moment at the same field (40\,T)\cite{Seb09.1} and thus
electrical transport signatures of a Kondo destruction would be expected.
Unfortunately, the presence of large fields may complicate such
analysis\cite{Bud05.2} and a careful modeling of the background contribution may
be needed.

The quantum critical behaviour discussed above occurs at the border of or inside
an AFM ordered phase. According to the phase diagrams in
Figs.\,\ref{phasediagrams} and \ref{pressure}, magnetism is more readily
suppressed by magnetic field or pressure than quadrupolar order. To the best of
our knowledge, no evidence for quadrupolar quantum criticality has been provided
to date in any Ce-based heavy fermion compound. A first indication thereof may
be the anomalous field dependence of the $A$ coefficient of the electrical
resistivity recently observed on a Ce$_3$Pd$_{20}$Si$_6$ single crystal in
magnetic fields applied along the $[100]$ direction.\cite{Mar14.1} For this
direction quadrupolar order is suppressed already at about
4\,T,\cite{Got09.1,Mit10.1,Ono13.1} and it is at this field where the anomaly in
$A(H)$ is seen.\cite{Mar14.1}

\section{Summary and outlook}

In this brief review, we have assembled information on Ce-based cubic heavy
fermion compounds that show low-lying, tunable phase transitions and are as
such candidates to further explore heavy fermion quantum criticality in the
three-dimensional limit.

A discriminating characteristic is the crystal field ground state of the
Ce$^{3+}$ $J=5/2$ multiplet. Quadrupolar order appears only if the $\Gamma_8$
quartet is the ground state (for at least one crystallographic site) or if there
is an important admixture (due to strong $c$--$f$ hybridization) of the
$\Gamma_8$ wave function in the ground state. Among the systems reviewed here,
CeIn$_3$ is the only one with a single (antiferromagnetic) phase transition. All
others show, in addition to a magnetic transition, a transition attributed to
higer multipoles, typically to quadrupoles.

To explore the three-dimensional limit, systems that retain cubic symmetry down
to zero temperature are needed. Two of the compounds, CeAg and
Ce$_3$Pd$_{20}$Ge$_6$, show structural transitions as the quadrupolar order sets
in. Interestingly, for both of them, this transition has been identified as
ferroquadrupolar. By contrast, at none of the antiferroquadrupolar transitions a
structural transition or distortion could be resolved.

Application of a magnetic field to a state with quadrupolar order typically
induces magnetic dipolar order on top of the ordered quadrupolar moments. As a
consequence, the quadrupolar phase transition temperature acquires field
dependence. This is seen for all examples discussed here as an initial
enhancement of the quadrupolar ordering temperature with magnetic field. In
addition, the field dependence of the quadrupolar ordering temperature is, in
general, anisotropic with respect to the direction along which the magnetic
field is applied (usually the directions $[001]$, $[110]$, and $[111]$ of the
cubic crystal structure are probed). This may, however, not be confused with a
breaking of the cubic symmetry of the crystal lattice, which is preserved as
long as no symmetry-lowering structural distortion occurs. Pure magnetic dipolar
order in cubic systems, on the other hand, is isotropic with respect to the
field direction, as experimentally observed for CeIn$_3$ with a $\Gamma_7$
ground state. This distinction may help to identify whether or not the magnetic
order in a given material is intimately coupled to quadrupolar order.
Details on how the different ordered moments interact with each other and with
the conduction electrons remain to be explored.

Finally, we would like to point out that the insulating sister compounds of
heavy fermion metals, fully-gapped Kondo insulators, are all cubic. The role of
quadrupolar (or higher multipolar) interactions in Kondo insulators is largely
unexplored, and so is the question whether or not these interactions are
important for the formation of topological Kondo insulators.\cite{Col13.1}

\section*{Acknowledgements}

We acknowledge fruitful discussions with Q.\ Si and A.\ Strydom, and financial
support from the European Research Council (ERC Advanced Grant No 227378) and
the Austrian Science Fund (FWF project I623-N16). JLJ acknowledges the FRC/URC
of the University of Johannesburg for funding of a Postdoctoral Fellowship under
joint supervision of SP and A.\ Strydom. We thank C.\ Marcenat for providing the plot in Fig.\,\ref{QCP}\,a.

\end{document}